\documentclass[11pt]{article}
\usepackage{epsfig,dsfont}

\begin{document}  
\sffamily

\thispagestyle{empty}
\vspace*{15mm}

\begin{center}

{\LARGE 
Center clusters in the Yang-Mills vacuum}
\vskip30mm
Christof Gattringer and Alexander Schmidt
\vskip8mm
Institut f\"ur Physik, Universit\"at Graz, \\
Universit\"atsplatz 5, 8010 Graz, Austria 

\end{center}
\vskip30mm

\begin{abstract}
Properties of local Polyakov loops for SU(2) and SU(3) lattice gauge 
theory at finite temperature are analyzed. We show that spatial clusters can 
be identified where the local Polyakov loops have values close to the same 
center element. 
For a suitable definition of these clusters the deconfinement 
transition can be characterized by the onset of percolation in one of the 
center sectors. The analysis is repeated for different resolution scales of 
the lattice and we argue that the center clusters have a continuum limit.
\end{abstract}
%\pacs{12.38.Aw, 11.15.Ha, 64.60.ah}
%\keywords{Lattice gauge theory, Confinement, Percolation}
%\maketitle

\newpage
\setcounter{page}{1}

\section{Introductory remarks}
 
Understanding the mechanisms that drive the transition to the deconfined
regime  is one of the great open problems of QCD. In particular with the
expected new  results from the experiments at the RHIC, LHC and GSI facilities
also the theoretical  side is challenged to contribute to our understanding of
confinement and the transition  to the deconfined phase.

Since phase transitions are non-perturbative phenomena, the applied methods
must be non-perturbative approaches. A particularly powerful technique is the
lattice formulation  of QCD, where numerical simulations have become reliable
quantitative tools of analysis.

An interesting idea, which is partly rooted in the lattice formulation, is the 
Svetitsky-Jaffe conjecture \cite{znbreaking} which links the deconfinement
transition of an SU($N$) gauge theory in $d+1$ dimensions to the magnetic
transition of a $d$-dimensional  spin system which is invariant under the
center group $\mathds{Z}_N$. The spins of the system are related
\cite{ploopeff} to the local Polyakov loops, which are static quark sources 
in the underlying gauge theory.

Having identified an effective spin system which describes SU($N$) gauge
theory  at the deconfinement transition, it is natural to ask whether one can
turn the  argument around and identify characteristic features of a spin system
directly in  the  corresponding gauge theory. Furthermore one may analyze
whether the gauge-spin relation holds only  at the critical temperature $T_c$
or also in a finite range of temperatures around $T_c$. 

A particular property of many discrete spin systems is the percolation of
suitably defined  clusters of spins at the magnetic transition. Since the spin
systems relevant for gauge theories have the discrete $\mathds{Z}_N$
invariance, one may expect to find some kind  of percolation phenomenon for
center degrees of freedom at the deconfinement transition of the gauge
theories.  Indeed, for the case of SU(2) lattice gauge theory studies of
percolation properties can be found in the literature
\cite{satz,fortunato,schmidt}, and more recently first results for SU(3)
\cite{gattringer} as well as full QCD \cite{borsanyi} were presented.

Establishing finite clusters below $T_c$ and percolating infinite clusters
above $T_c$ gives rise to a tempting interpretation of the deconfinement
transition: The size of finite  clusters in the confining phase might be
related to the maximal distance one can place a quark and an anti-quark source
such that they still have a non-vanishing vacuum expectation value. For larger
distances the two sources always end up in different clusters and average to
zero independently. Above $T_c$ there exists an infinite cluster and with a
finite probability the two sources are correlated also at arbitrary distances
such that they can move freely.       

However, the above sketched picture hinges crucially on the scaling properties
of the center clusters -- a question that so far has not been addressed in the
literature. A spin system has an intrinsic scale: The lattice constant of the
underlying grid. In lattice gauge theory the situation is different: There one
is interested in studying the system for finer and finer lattices in order to
learn about the continuum limit. For the percolation picture this implies that
when measured in lattice units, the clusters have to be larger for finer
lattices. Only then the size of the cluster in physical units, e.g., the
diameter of the cluster multiplied with the lattice constant in fm can approach
a finite value and can be assigned a  physical meaning. If no such scaling 
behavior can be established the clusters are merely lattice artifacts. 

In this article we compare for SU(3) and SU(2) lattice gauge theory the
evidence for  the existence of center clusters and their percolation at $T_c$.
Particular focus is put on the analysis of the scaling behavior of the
clusters. We study the flow of the cluster parameters as a function of the
lattice spacing and demonstrate that a continuum limit for the cluster 
picture is plausible. 

\section{Conventions and setting of our analysis}

In our analysis we explore pure SU(3) and SU(2) lattice gauge theory at 
temperatures below and above the deconfinement transition. The basic 
observable we analyze is the local Polyakov loop $L(\vec{x})$ defined as  
\begin{equation}
L(\vec{x}) \, = \, 
\mbox{Tr} \prod_{t=1}^{N_t} U_4(\vec{x},t) \; . 
\label{Ploopdef}
\end{equation}
$L(\vec{x})$ is the ordered product of the SU(3) or SU(2) valued temporal
gauge  variables $U_4(\vec{x},t)$ at a fixed spatial position $\vec{x}$, where
$N_t$ is the number of lattice points in time direction and $\mbox{Tr}$
denotes the trace over color indices.  The loop $L(\vec{x})$ thus is a gauge
transporter that closes around  compactified time. Often also the spatially
averaged loop $P$ is considered, which we define as
\begin{equation}
P \; =  \; \frac{1}{V} \, \sum_{\vec{x}} L(\vec{x})\; ,
\label{averloop}
\end{equation}
where $V$ is the spatial volume.  Due to translational
invariance $P$ and $L(\vec{x})$ have the same vacuum expectation value. 

The Polyakov loop corresponds to a static quark source and its vacuum
expectation value is (after a suitable renormalization)  related to the free
energy $F_q$ of a single quark, $\langle L(\vec{x}) \rangle = \langle P
\rangle  \propto \exp(-F_q/T)$, where $T$ is the temperature  (the Boltzmann
constant is set to 1 in our units). Below the critical temperature $T_c$
quarks are confined and $F_q$ is infinite, implying $\langle L(\vec{x})
\rangle = \langle P \rangle = 0$. The transition from the confined to the
deconfined phase is of first order for the case of SU(3), while it is second
order for SU(2) gauge theory.

The deconfinement transition of pure Yang-Mills theory may also be interpreted
as the spontaneous breaking of center symmetry. For SU(3) the elements $z$ of
the center group  $\mathds{Z}_3$ are a set of three phases, $z \in \{ 1 , e^{i
2\pi/3} , e^{-i 2\pi/3} \}$, while for SU(2) we have the center group
$\mathds{Z}_2$ with  $z \in \{1, -1\}$. In a center transformation all temporal
links in a fixed time slice are multiplied with an element $z$ of the center
group. While the action and the path integral measure are invariant under a
center transformation, the local and averaged Polyakov loops transform
non-trivially as
\begin{equation}
L(\vec{x}) \; \longrightarrow \; z \, L(\vec{x}) \qquad \mbox{and} \qquad P \;
\longrightarrow \; z \, P \; .
\end{equation}
The non-vanishing expectation value  $\langle L(\vec{x}) \rangle = \langle P
\rangle \neq 0$, which we find above $T_c$, thus signals the spontaneous 
breaking of the center symmetry. 

In our study we analyze the behavior of the local Polyakov loops $L(\vec{x})$
using quenched SU(3) and SU(2) configurations at finite temperature. For both
gauge groups we use the L\"uscher-Weisz gauge action \cite{LuWe} on lattices
with different sizes ranging from $20^3 \times 6$ to $40^3 \times 12$. For
SU(3) the lattice constant $a$ was determined in \cite{scale} using the Sommer
parameter, while for SU(2) we express dimensionful quantities using suitable
powers of the string tension $\sigma$ at zero temperature. For SU(3) the 
critical temperature
determined in \cite{TCdet} is used, for SU(2) it was determined using the 
Polyakov loop susceptibility \cite{schmidt}. All errors we show are statistical
errors determined with single elimination Jackknife. 

\section{Distribution properties of local Polyakov loops}

We begin our analysis with studying the distribution properties of the local
Polyakov loops $L(\vec{x})$ defined in (\ref{Ploopdef}). While $L(\vec{x})$ is
a real number for the gauge group SU(2) it is complex for SU(3). For the latter
case we decompose the local Polyakov loops into modulus and phase,
\begin{equation}
L(\vec{x}) \; = \; \rho(\vec{x}) \, e^{ i \varphi(\vec{x}) } \; .
\end{equation}
In \cite{gattringer} it was demonstrated that the distribution of the modulus
$\rho(\vec{x})$ is a rather unspectacular quantity. It is very well described
by the Haar measure distribution $P(\rho) = \int dU \delta(\rho - | \mbox{Tr} U
|)$, where $dU$ denotes the integration according to Haar measure over SU(3)
group elements $U$. In particular the distribution of the modulus is almost
entirely insensitive to temperature, lattice volume and resolution scale
\cite{gattringer}. Thus
we conclude that the first order transition of SU(3) lattice gauge theory 
into the
deconfined phase is not driven by a change in the distribution of the modulus. 

\begin{figure}[t]
\begin{center}
\includegraphics[width=12.5cm,clip]{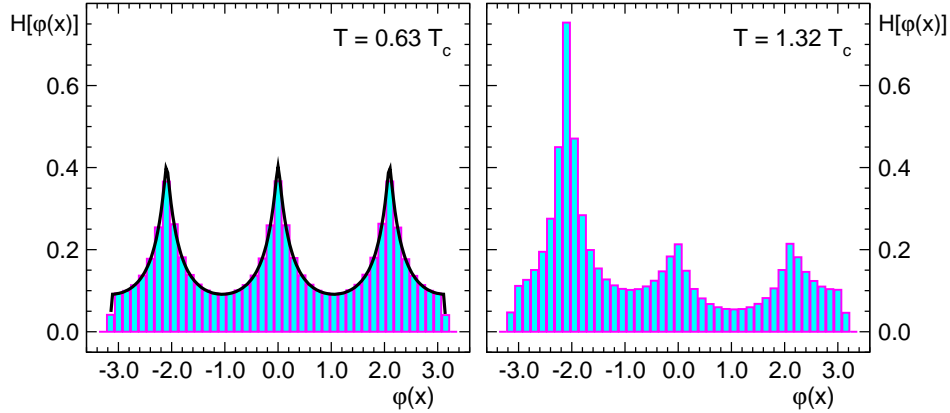} 
\end{center}
\caption{Histograms for the distribution of the phases of local Polyakov loops
for the case of SU(3). The results are from our $40^3 \times 6$ SU(3) ensembles
and we compare two temperatures, $T=0.63 \, T_c$ (lhs.\ plot) and $T = 1.32 \,
T_c$ (rhs.). The curve superimposed on the low temperature histograms is the
Haar measure distribution  $P(\varphi) = \int dU \delta( \varphi - \,
\mbox{arg}\, \mbox{Tr} \, U )$. The high temperature results are for the sector
of gauge configurations characterized by the SU(3) center element $z =
e^{-i2\pi/3}$. 
\label{ploophistoSU3}}
\end{figure}

The situation is different for the distribution of the phase
$\varphi(\vec{x})$, where indeed we observe a strong change in the distribution
as one crosses into the deconfined phase. This is illustrated in
Fig.~\ref{ploophistoSU3}, where we show histograms $H[\varphi(\vec{x})]$ for
the distribution of the phases $\varphi(\vec{x})$  comparing the two
temperatures $T = 0.63 \, T_c$ and $T = 1.32 \, T_c$. Below $T_c$ (lhs.\ plot),
the distribution shows three pronounced peaks located at the center phases $-2
\pi/3, 0$ and $+2\pi/3$. The whole $T = 0.63 \, T_c$ histogram perfectly
matches the Haar measure distribution  $P(\varphi) = \int dU \delta( \varphi -
\, \mbox{arg} \, \mbox{Tr} \, U )$ which we superimpose as a full curve on the
histograms in the lhs.\ plot of Fig.~\ref{ploophistoSU3}.

Above $T_c$ (rhs.\ plot) the distribution changes: One of the peaks has grown,
while the other two peaks have shrunk. We stress at this point, that the
selection which peak becomes enhanced is a matter of spontaneous symmetry
breaking. In the rhs.\ plot of  Fig.~\ref{ploophistoSU3} the system has chosen
the sector which is characterized by a phase $e^{-i 2\pi/3}$ of the spatially
averaged Polyakov loop $P$ (compare Eq.\ (\ref{averloop})). In case one of the
other two center sectors is selected, the whole distribution in the rhs.\ plot
is shifted periodically 
by $\pm 2\pi/3$. It is interesting to note, that also above $T_c$ we
still see the subdominant  peaks which correspond to the center sectors that
were not selected in the act of spontaneous symmetry breaking. This already
hints at the possibility that spatial  bubbles with phases $\varphi(\vec{x})$
near the subdominant center phases might exist also above $T_c$.

\begin{figure}[t]
\begin{center}
\includegraphics[width=12.5cm,clip]{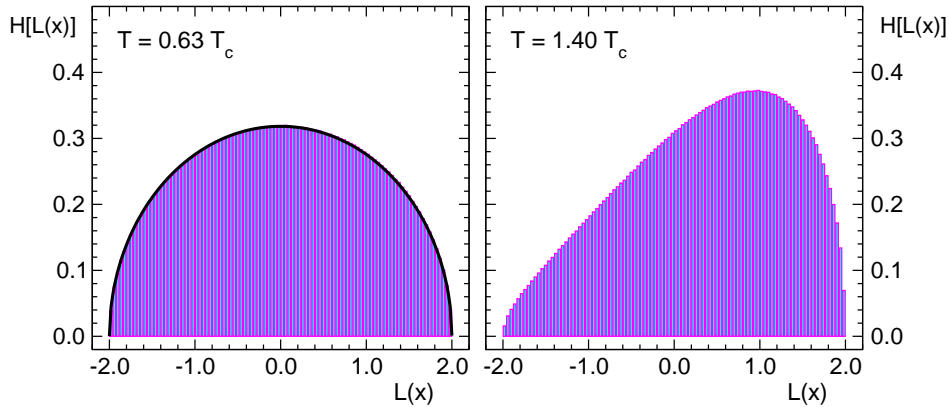} 
\end{center}
\caption{Histograms for the distribution of the local Polyakov loops for the
case of SU(2). The results are from our $40^3 \times 6$ SU(2) ensembles and we
compare two temperatures, $T=0.63 \, T_c$ (lhs.\ plot) and  $T = 1.40 \, T_c$
(rhs.). The curve superimposed on the low temperature histograms is the Haar
measure distribution  $P(L) = \int dU \delta( L - \, \mbox{Tr} \, U)$. The high
temperature results are for the sector of gauge configurations characterized by
the SU(2) center element $z = +1$. 
\label{ploophistoSU2}}
\end{figure}

Let us now come to the case of SU(2). There the local Polyakov loops
$L(\vec{x})$ are real and we can directly look at their distribution. In
Fig.~\ref{ploophistoSU2} we show histograms $H[L(\vec{x})]$ for the
distribution of $L(\vec{x})$, again comparing two temperatures, $T = 0.63 \,
T_c$ (lhs.\ plot) and $T = 1.4 \, T_c$ (rhs.). Similar to the SU(3) case we
find that below $T_c$ the distribution very closely follows the Haar measure
distribution $P(L) = \int dU \delta( L - \, \mbox{Tr} \, U)$, which we
superimpose as a full curve in the corresponding plot. Above $T_c$ we observe a
deformation of the distribution favoring positive values. However, again we
stress that this is a manifestation of spontaneous symmetry breaking, since
here we use high temperature configurations which are characterized by a
positive value of the spatially averaged Polyakov loop $P$. In case the system
spontaneously selects the sector of the other center element $z = -1$, which is
characterized by negative  values of $P$ above $T_c$, the distribution in the
rhs.\ plot of Fig.~\ref{ploophistoSU2} would display its peak at negative
values of $L(\vec{x})$.

The spontaneous symmetry breaking, which leads to a non-vanishing expectation
value of the Polyakov loop above $T_c$, can now be directly related to the
distributions of the local Polyakov loops $L(\vec{x})$. In particular the
relative distributions in the three (two) center sectors for gauge group SU(3)
(gauge group SU(2)) drive the change of the expectation value of the Polyakov
loop. The three sectors of SU(3) are here defined by the three intervals
$[-\pi, -\pi/3]$, $[-\pi/3,+\pi/3]$ and $[\pi/3,\pi]$, the phases
$\varphi(\vec{x})$ can fall into. The two sectors for the case of SU(2) are
distinguished by the sign of $L(\vec{x})$. 

\begin{figure}[t]
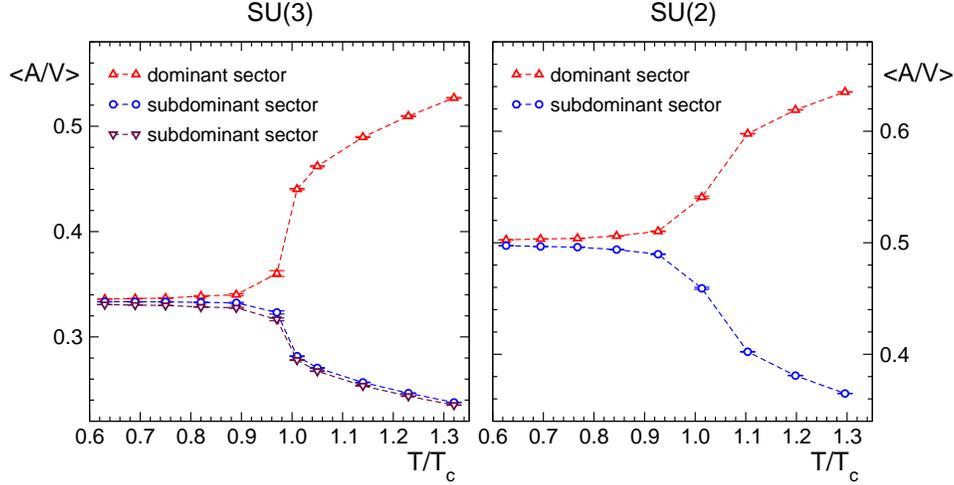

\begin{center}
\includegraphics[height=6.5cm,clip]{abundance_vs_ttc_40x6.eps}
\hspace{-2mm}
\includegraphics[height=6.5cm,clip]{abundance_vs_ttc_SU2_40x6.eps}
\end{center}
\caption{Abundance $A$ of sites in each sector 
normalized by the spatial volume $V$ as
a function of temperature. The results are for 
our $40^3 \times 6$ lattices and we
compare  the case of SU(3) (lhs.\ plot) with SU(2) gauge theory (rhs.).    
\label{abundance}}
\end{figure}

Below $T_c$ all three (two) sectors are populated with an equal number of 
sites $\vec{x}$. When all center sectors are equally  populated, the Polyakov
loop expectation value vanishes as the center elements add up to zero, i.e., $1
+ e^{i2\pi/3} + e^{-i2\pi/3} = 0$ for SU(3), and $1 \, + \,(-1) = 0$ for SU(2).
Above $T_c$ one of the center sectors is more populated and thus the
contributions do no longer add up to zero. In Fig.~\ref{abundance} we show the
abundance $A$ of sites in each sector normalized by the spatial volume $V$  as
a function of temperature and compare SU(3) (lhs.\ plot) and SU(2) (rhs.).  For
low temperature each of the sectors is for the case of SU(3) populated with
roughly a third of the lattice sites (half of the lattice sites for SU(2)).
Near $T_c$ one of the sectors starts to increase the abundance of sites, while
the other sectors become depleted. It is interesting to note that for the case
of SU(3), where the transition is first order, the curves show a rather sudden
change near $T_c$, while for SU(2), where the transition is of second order,
the behavior is smoother (as expected).

\section{Center clusters}

In the previous section we have studied the distribution 
of the local Polyakov loops $L(\vec{x})$. When analyzing the abundance of
sites assigned to the different center sectors we found that near $T_c$ one of
the sectors starts to become more populated, thus giving rise to the
non-vanishing expectation value of the Polyakov loop. In this section we now
analyze the connectedness properties of sites $\vec{x}, \vec{y}$ where the
values of the local loops  $L(\vec{x}), L(\vec{y})$ fall in the same center
sector. For that purpose we assign sector numbers $n(\vec{x})$ to the sites
$\vec{x}$. For the case of SU(3) the sector numbers can have three values
$n(\vec{x}) \in \{-1,0,+1\}$ assigned according to 
\begin{equation}
n(\vec{x}) \; = \; \left\{ \begin{array}{rl}
-1 & \; \mbox{for} \;\;\;\; \varphi(\vec{x}) \, \in \, 
[\,-\pi + \delta \; , \; -\pi/3 - \delta \, ] \; ,\\
0 & \; \mbox{for} \;\;\;\; \varphi(\vec{x}) \, \in \, 
[\,-\pi/3 + \delta \, , \, \pi/3 - \delta \, ] \; ,\\
+1 & \; \mbox{for} \;\;\;\; \varphi(\vec{x}) \, \in \, 
[\,\pi/3 + \delta \, , \, \pi - \delta \,] \; , 
\end{array} \right. 
\label{sectornumbers}
\end{equation}
while for SU(2) we have
two possibilities, $n(\vec{x}) \in \{-1,+1\}$, with
\begin{equation}
n(\vec{x}) \; = \; \left\{ \begin{array}{rl}
-1 & \; \mbox{for} \;\;\;\; L(\vec{x}) \; \leq \; - \, \delta \; ,\\
+1 & \; \mbox{for} \;\;\;\; L(\vec{x}) \; \geq \; + \, \delta \; .\\ 
\end{array} \right. 
\label{sectornumberssu2}
\end{equation}
If the phase $\varphi(\vec{x})$ (the local loop $L(\vec{x})$ for SU(2)) is not
contained in one of the intervals, no sector number is assigned to the site
$\vec{x}$, which then is no longer taken into account in the subsequent
analysis. We can now define center clusters by assigning two neighboring sites
$\vec{x}$ and $\vec{y}$ to the same cluster if $n(\vec{x}) = n(\vec{y})$.

Let us comment on the role which the parameter $\delta$ plays in
our construction of the center clusters. The parameter $\delta$ allows one to
cut those lattice sites where the corresponding local Polyakov loop does not
clearly lean towards one of the center elements. For the case of SU(3) sites
$\vec{x}$ with phases near the minima of the distributions shown in
Fig.~\ref{ploophistoSU3} are cut, while for SU(2) the cut leads to a removal of
sites where $L(\vec{x})$ is close to zero. In order to have a more accessible
definition of how much we cut, instead of quoting a value of the parameter
$\delta$, from now on we express the cut in percent of sites that are not
assigned a sector number $n(\vec{x})$, i.e., the percentage of sites that are 
removed from the analysis. To allow for a comparison of the values $\delta$ 
and the 
cut in \%, we list the corresponding numbers for $30^3 \times 6$ at three
temperatures in Table~1. The percentage of points that are cut 
for a given $\delta$ 
shows a mild temperature dependence, and when refering to these numbers in the 
subsequent text we quote an 
average value. The volume dependence is negligible. 

\begin{table}[t]
\begin{center}
\begin{tabular}{c|c|c|c} 
$\delta$ & sites cut, $0.69 \, T_c$ & sites cut, $1.01 \, T_c 
$ &sites cut, $1.22 \, T_c$ \\
\hline
$0.0 $ & 0.00 \% & 0.00 \% & 0.00 \%\\
$0.1 \, \pi/3 $ &  5.76 \% &  5.67 \% &  5.57 \% \\
$0.2 \, \pi/3 $ & 11.67 \% & 11.51 \% & 11.28 \% \\
$0.3 \, \pi/3 $ & 17.85 \% & 17.61 \% & 17.25 \% \\
$0.4 \, \pi/3 $ & 24.48 \% & 24.21 \% & 23.72 \% \\
$0.5 \, \pi/3 $ & 31.83 \% & 31.46 \% & 30.86 \% \\
$0.6 \, \pi/3 $ & 40.18 \% & 39.76 \% & 39.06 \% \\
\end{tabular} 
\end{center}
\caption{Comparison of the cut parameter values $\delta$ and the fraction of
cut sites in $\%$. We show the SU(3) results for our $30^3\times6$ lattices and 
compare different temperatures.}
\end{table}

\begin{figure}[t]
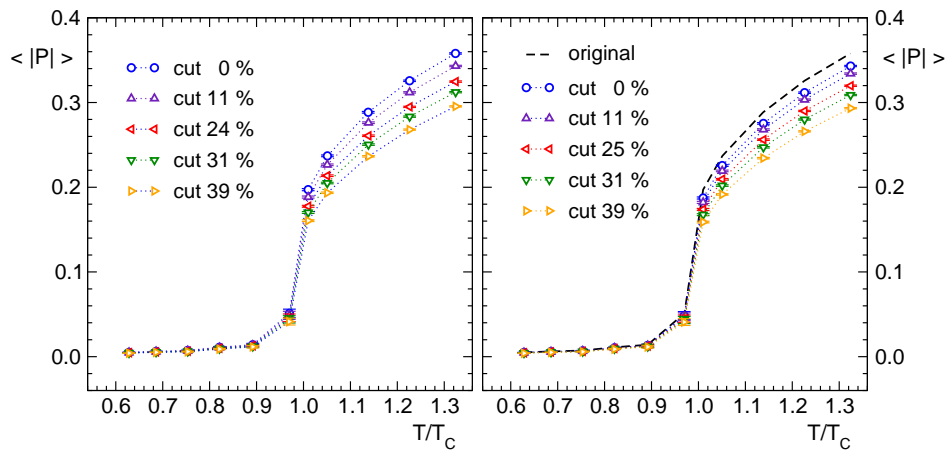

\begin{center}
\includegraphics[height=5.9cm,clip]{Ploop_cut.eps}
\hspace{-2mm}
\includegraphics[height=5.9cm,clip]{Ploop_cut3.eps}
\end{center}
\caption{Lhs.\ plot: Expectation value of the modulus of the Polyakov loop when
only those sites are taken into account that survive the cut defined in
(\ref{sectornumbers}). Rhs.: Expectation value of the Polyakov loop when using
the nearest center  elements instead of the local loop $L(\vec{x})$. Both
figures are for SU(3) on $40^3 \times 6$ lattices and we compare different
values of the cut.   
\label{ploopcut}}
\end{figure}

Introducing such a cut for the analysis of cluster properties of course 
immediately raises the question whether such a cut does not destroy the physics
one wants to study. In order to address this question we show in the lhs.\ plot
of Fig.~\ref{ploopcut} the results for the Polyakov loop as a function of
temperature for different amounts of lattice sites removed when turning on
$\delta$ (we show results for SU(3) on $40^3 \times 6$). It is obvious that
even a cut  of almost 40 \% of lattice points leads to only a small reduction
of the expectation value  of the Polyakov loop, indicating that the cut indeed
removes only the rather unimportant fluctuations between the center values. One
can even go one step further and replace the local loop $L(\vec{x})$ by the
nearest center element. The result is shown in the rhs.\ plot of
Fig.~\ref{ploopcut}, again comparing different values for the cut. It is
obvious that the center elements alone are sufficient to reproduce most of the
Polyakov loop expectation value. Repeating the same analysis for SU(2) leads to
equivalent results \cite{schmidt}.

Having convinced ourselves that the cut does not destroy the physics we want to
analyze, let us add a few more comments on the role of the cluster parameter
$\delta$. It is obvious that a small value of $\delta$ will produce denser
clusters as more points are available for forming the clusters. As one increases
$\delta$ the clusters become thinner and smaller. One of the motivations of this
analysis is to study a possible characterization of the deconfinement
transition as a  percolation phenomenon. This is an idea that has been widely
explored in the context of spin systems, where for many models the magnetic
transition may be characterized by the percolation of suitably defined
clusters. For these systems it is well known that a naive cluster definition,
where neighboring sites with equal spin values are assigned to the same
cluster, gives rise to clusters that are too dense, such that the percolation-
and the magnetic transitions do not  coincide. Only if the clusters are
"thinned out" the two critical temperatures will agree. In particular one may
use the Fortuin-Kasteleyn cluster construction \cite{fkclusters}, which is now
understood to give the correct cluster description of the magnetic transition
in Potts models \cite{coniglio}. The parameter $\delta$ which we introduce in
our cluster construction plays exactly the same role as the more educated
cluster definitions in spin systems, such as the Fortuin-Kasteleyn
construction. As a matter of fact the introduction of a free parameter for
controlling the cluster density, similar to our prescription, has been discussed
in the literature \cite{pbond}. 

There is, however, an important difference between the analysis of percolation
in spin systems and in lattice gauge theory. While in the former case there is
a fixed scale, the lattice spacing of the ferromagnet one studies, in the
analysis of lattice gauge theories one is interested in performing the continuum
limit, i.e., the limit of vanishing lattice constant $a \, \rightarrow \, 0$. 
The cluster picture only  has a chance for a reasonable continuum limit, if the
cluster diameter in lattice units diverges as one approaches $a = 0$. Only
in that case the cluster diameter in physical units, which is obtained as the
product of the diameter in lattice units with the lattice constant $a$,
can have a finite limit. As we will demonstrate in the next section, our cluster
definition with the parameter $\delta$ is suitable for such an analysis of a
possible continuum limit.

\begin{figure}[t]
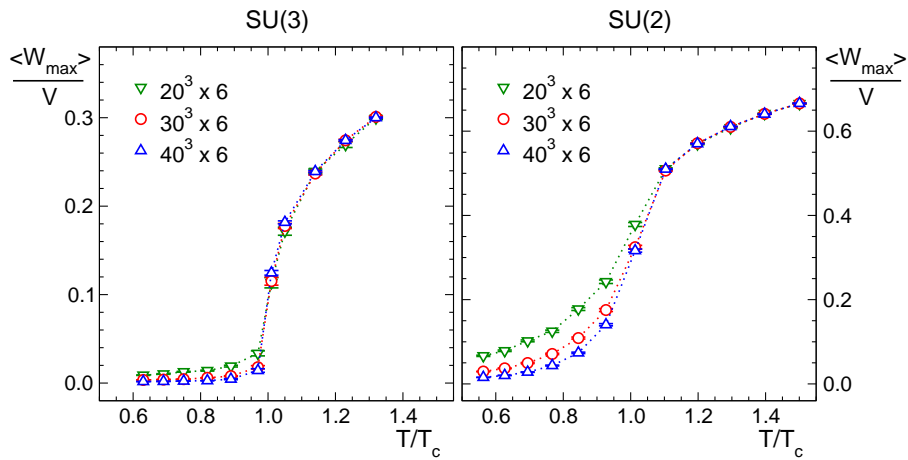

\begin{center}
\includegraphics[height=6.00cm,clip]{weight_maxcluster_SU3.eps}
\hspace{-2mm}
\includegraphics[height=6.00cm,clip]{weight_maxcluster_SU2.eps}
\end{center}
\caption{Weight $W_{max}$ of the largest cluster normalized by the volume as a
function of temperature. We compare the case of SU(3) (lhs.\ plot) with SU(2)
gauge theory (rhs.) and display data for three different volumes. The cuts we
use are 39 \% of sites for SU(3) and 48 \% for SU(2).   
\label{weight_maxcluster}}
\end{figure}

Having discussed our cluster construction and the role of the parameter
$\delta$, we now have a first look at a possible percolation behavior near
$T_c$. For that purpose we analyze the weight $W_{max}$ of the largest
cluster,  i.e., the number of sites in the largest cluster, as a function of
temperature. In Fig.~\ref{weight_maxcluster} we display the expectation value
of $W_{max}$ normalized by the spatial volume $V$ as a function of $T$. We
show the results for SU(3) in the lhs.\ plot (for a cut of 39 \%) and the
results for SU(2) on the rhs.\ (for a cut of 48 \%). In both cases we compare
three different volumes. 

Below $T_c$ we find that $\langle W_{max}\rangle/V$ depends on the spatial
volume $V$. The observation that for a fixed lattice constant
$\langle W_{max}\rangle/V$ decreases with increasing volume $V$ suggests that
below $T_c$ the clusters have a finite maximal size 
$\langle W_{max}\rangle \sim
const$. 
Above $T_c$ the volume dependence is gone and the
largest cluster fills a certain fraction of the total volume with a finite
density that keeps growing as one further increases the temperature.  This
absence of a volume dependence indicates that the clusters are percolating
above $T_c$. For a direct analysis of the percolation probability as function
of temperature see \cite{schmidt,gattringer,borsanyi}.

\section{Continuum limit of the center clusters and the percolation picture}

In the previous section we have demonstrated that for suitably defined  center
clusters percolation sets in at $T_c$. So far the  free parameter $\delta$
which we introduced in our cluster construction was chosen arbitrarily. Now we
change our approach and use a physical scale to set the cluster parameter
$\delta$. The scale we use is the diameter of the  clusters in physical units.
With such a prescription we can compare properties of center clusters on
lattices with different resolution $a$. 

In order to define the diameter of the clusters we consider the correlation
functions of points within the same cluster, belonging to the center sector
characterized by the sector number $n$,
\begin{equation}
C_n(r) \; = \; \frac{1}{3 N_n} \sum_{\vec{x}} \sum_{\mu = 1}^3 \; \Delta_n (
\vec{x}, \vec{x} + r \hat{\mu} ) \; .
\label{correlatordef}
\end{equation}
The sector number $n$ can have the three values $n \in \{-1,0,+1\}$ for SU(3),
while for SU(2) we have $n \in \{-1,+1\}$. By $N_n$ we denote the total number
of sites $\vec{x}$ with section number $n(\vec{x}) = n$. The first sum runs
over all lattice sites, the second one over all three spatial directions $\mu$
and by $\hat{\mu}$ we denote the corresponding unit vector. The parameter $r$
assumes values $r = 0, 1, 2, ...\, $, i.e., in (\ref{correlatordef}) we consider
the correlation along the coordinate axes. The function $\Delta_n (
\vec{x}, \vec{y} )$ is defined through 
\begin{equation}
\Delta_n (
\vec{x}, \vec{y} ) \, = \left\{ \begin{array}{rl}
\!\!1 & \mbox{if} \; \vec{x} \; \mbox{and} \; \vec{y} \; \mbox{are in the same
cluster, and this cluster is type} \; n, \\
\!\!0 & \mbox{else} \; .
\end{array}
\right.
\end{equation}
\begin{figure}[t]
\begin{center}
\includegraphics[height=5.7cm,clip]{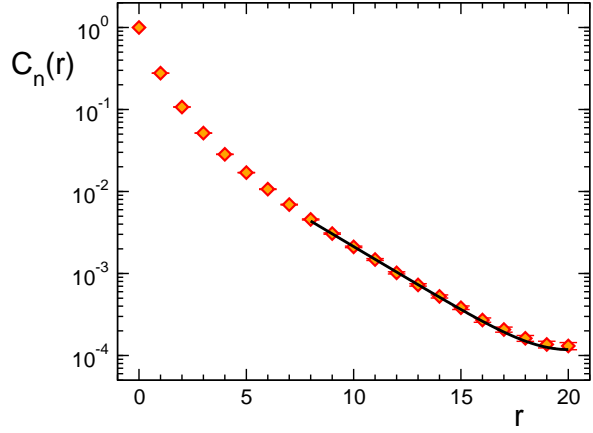}
\end{center}
\caption{Example for the cluster correlation function $C_n(r)$ as a function of
$r$ (symbols), and the corresponding fit with the functional form of
(\ref{fitfunct}) (full curve). The data are for SU(3), $40^3 \times 12$ 
at $T = 0.64 \, T_c$ with a cut of 18 \%.    
\label{correlator}}
\end{figure}
Up to some correction at short distances, these correlation functions decay
exponentially in $r$. Similar to euclidean correlators in lattice spectroscopy,
for sufficiently large $r$ we fit them with the function 
\begin{equation}
C_n(r) \; \sim \; A \, \cosh((r - L/2)/\rho) \; ,
\label{fitfunct}
\end{equation} 
where $A$ and $\rho$ are two real fit parameters and $L$ is the number of
lattice points in the spatial direction. As always, for correlation functions
in a finite volume the hyperbolic cosine appears due to the periodic 
spatial boundary
conditions which we use.  An example of the correlation function and the
corresponding fit is shown in  Fig.~\ref{correlator} for the case of SU(3).

The value of $d_{lat} = 2 \rho$ is now used as the diameter in lattice units. It
is converted to the diameter in physical units $d_{phys}$ by multiplication with
the lattice constant $a$, such that we obtain the diameter $d_{phys} = a \,
d_{lat}$ in fm (or in units of
the string tension for SU(2)). We now use the physical scale $d_{phys}$ to set
the value for the cluster parameter $\delta$. 
For example we may decide to analyze
clusters that have a fixed diameter of $d_{phys} = 0.5$ fm defined at some
fixed temperature, e.g., $T = 0.63 \, T_c$. Then, working on the $T = 0.63\,
T_c$ ensembles, we adjust the parameter $\delta$ such, that our diameter
$d_{phys}$ has the desired value of $d_{phys} = 0.5$ fm. This procedure can now
be repeated on lattices with different lattice constant $a$ and we thus can
study scaling effects and the continuum limit.

\begin{figure}[t]
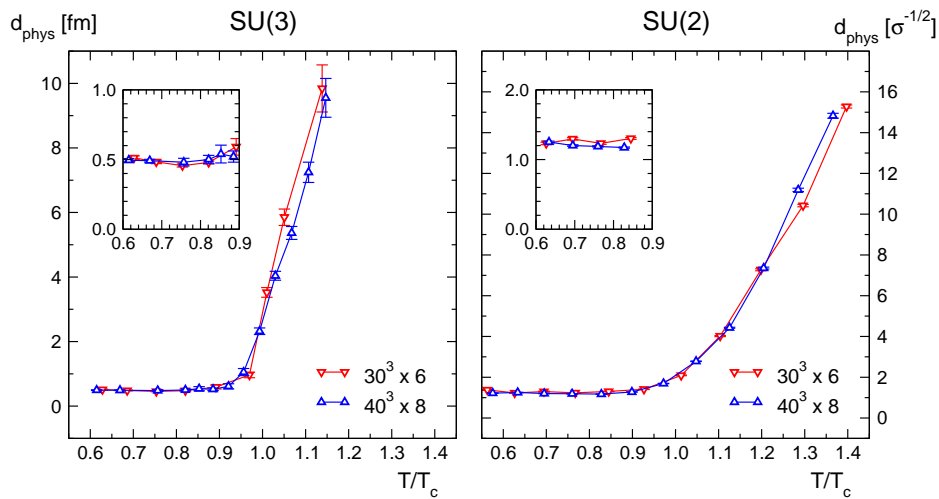

\begin{center}
\includegraphics[height=6.53cm,clip]{scaling_diameter.eps}
\hspace{-0mm}
\includegraphics[height=6.53cm,clip]{scaling_diameter_SU2.eps}
\end{center}
\caption{Diameter $d_{phys}$ of the clusters in physical units as 
a function of temperature for the
case of SU(3) (lhs. plot) and SU(2) (rhs.). The inserts are a zoom into the low
temperature data. In both cases we compare the
results for two different values of the lattice constant $a$. 
\label{physical_size}}
\end{figure}

Based on that physical definition of the center clusters we now address the
question how the average diameter of the clusters in physical units behaves as
a function of temperature. For that purpose we proceed as described in the last
paragraph and for a given lattice constant $a$ we adjust the cluster parameter
$\delta$ such that the clusters have an average diameter of $d_{phys} = 0.5$ fm
(which is a reasonable mesonic scale for very heavy quarks). We then keep this
value of the cluster parameter $\delta$ fixed and change the temperature, to
study how the physical cluster diameter changes with $T$. Keeping $\delta$ fixed
simply means that we work with the same cluster construction for all temperatures, 
i.e., the percentage of points we cut is almost the same for all $T$
(compare Table 1). The result of this
analysis is presented in Fig.~\ref{physical_size} where we show the cluster diameter
$d_{phys}$ in fm for the case of SU(3) (lhs.\ plot), and for SU(2) in units of
the string tension (rhs.), as a function of temperature. For both cases we find
that below $T_c$ the cluster diameter in physical units remains essentially
constant  at the value of $d_{phys}$ that we dialed in at $T = 0.63 \, T_c$. At
the critical temperature the cluster diameter starts to rise quickly indicating
that the clusters start to percolate. We stress at this point that on an
infinite lattice the curve for the diameter would jump to infinite slope at
$T_c$. On a finite lattice with $L$ lattice points in the spatial directions,
the cluster diameter  we can determine using the correlator $C_n(r)$ is always
bounded by the  finite spatial box size, giving rise to a finite slope in
Fig.~\ref{physical_size} above $T_c$. Comparing different values of the box
size $a L$ in physical units, we found that the results for  $d_{phys}$ 
at a fixed value of $T \, (> T_c)$ increase with $L$. We stress at this point
that while it is obvious that at fixed $T$ the values for $d_{phys}$ increase
with $L$, it is not a priori clear that for fixed $L$ the value of
$d_{phys}$ is an
increasing function of $T$, because we drive the temperature by changing the
lattice constant $a$, such that also the physical volume shrinks. The fact
that  above $T_c$ $d_{phys}$ keeps growing with $T$ is due to an increase of
the density of the percolationg cluster which can, e.g., be seen from Fig.~5,
which shows that the number of sites in the largest cluster keeps growing 
above $T_c$.

The crucial check for our analysis of the cluster diameter in physical units is
the comparison of the results obtained on ensembles with different lattice
spacing $a$. For that purpose in Fig.~\ref{physical_size} we compare the curves 
for the cluster diameters
on $30^3 \times 6$ and $40^3 \times 8$ lattices. In both cases we followed the
above described procedure, adjusted the diameter to a fixed physical  value
$d_{phys}$ at $T = 0.63 \, T_c$ and kept the corresponding cluster parameter
$\delta$ for all temperatures. If the cluster picture and the percolation
properties we found have a meaning in physical units, the two curves should
fall on top of each other. The plots show that this is the case for
both SU(3) and SU(2).

Having analyzed the percolation picture in physical units and finding universal
behavior when comparing different lattice spacings, let us now turn to another
conundrum related to the continuum limit of the center cluster picture.
Fig.~\ref{physical_size} suggests that the center clusters have a fixed physical
diameter below $T_c$ (at least in the range of temperatures we studied), and
that this behavior can be seen on lattices with different resolution $a$.  
This poses the questions how below $T_c$ the physical diameter 
\begin{equation}
d_{phys} \; = \; a \, d_{lat}  \; ,
\label{dphys}
\end{equation}
can remain finite in the limit $a \rightarrow 0$. Equation (\ref{dphys})
implies that for the construction of the continuum limit the cluster diameter 
in lattice units $d_{lat}$  has to diverge, in other words the clusters must
become infinite in that limit. 

\begin{figure}[t]
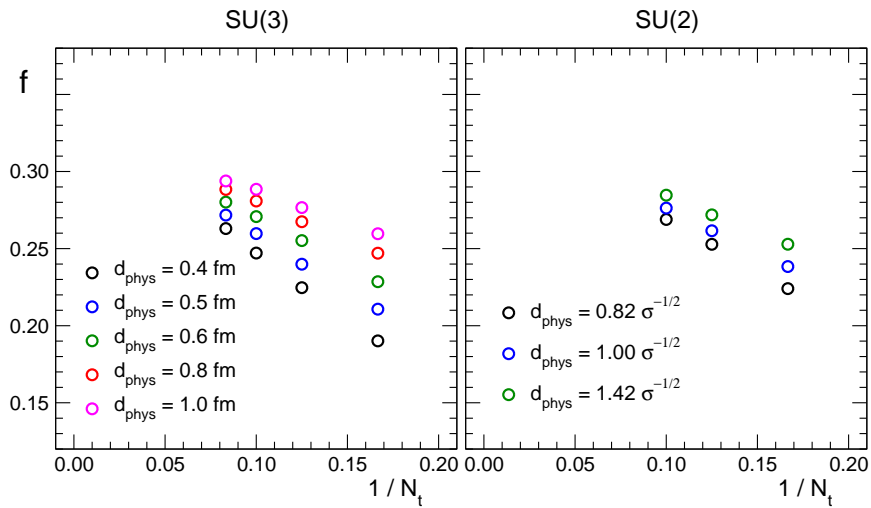

\begin{center}
\hspace*{3mm}
\includegraphics[height=6.65cm,clip]{percentageavailable_vs_Ntinv.eps}
\hspace{-2mm}
\includegraphics[height=6.65cm,clip]{percentageavailable_vs_Ntinv_SU2.eps}
\end{center}
\caption{The fraction $f$ of points available for clusters per center sector for the case of
SU(3) (lhs.\ plot) and SU(2) (rhs.). We compare different physical cluster diameters,
which for the case of SU(3) we give in fm, while for SU(2) the cluster diameter is
given in units of the string tension.   
\label{fractionavailable}}
\end{figure}

We address this question by comparing clusters
of the same physical size on lattices with different resolution $a$. This
comparison is done at the fixed temperature $T = 0.63 \, T_c$. For each value of
$a$ we set the cluster parameter $\delta$ 
such that the physical cluster diameter $d_{phys}$ has the desired size, e.g.,
$d_{phys} = 0.5$ fm. For
different values of the lattice spacing $a$ different values of $\delta$ are
necessary to obtain the desired $d_{phys}$: For finer lattices we need larger
clusters and thus can cut only fewer points than for coarse lattices. By 
$N_{cut}$ we denote the number of lattice sites that are removed when
constructing the clusters according to Eqs.\ (\ref{sectornumbers}) and
(\ref{sectornumberssu2}).  We measure
the influence of the parameter $\delta$ by defining the following fraction
\begin{equation}
f \; = \; \frac{V \, - \, N_{cut}}{V \, N_c} \; .
\end{equation}
This fraction measures how many sites are available  (= the numerator $V -
N_{cut}$), per volume (factor $1/V$), per center sector (factor $1/N_c$, where
$N_c$ is the number of center elements). 
In other words $f$ measures the fraction of
sites available for clusters in a given center sector. In
Fig.~\ref{fractionavailable} we show the flow of $f$ with $1/N_t$. The
continuum limit is reached for  $1/N_t \rightarrow 0$. We compare the results
for different values of $d_{phys}$, which for the case of SU(3) we measure in
fm (lhs.\ plot), while for SU(2) (rhs.) we express $d_{phys}$ in units of the
string tension.
 
It is obvious from the plots that for a given $d_{phys}$ the values for the
available fraction $f$ at different $1/N_t$ fall on almost perfect straight
lines. Moreover, all the lines extrapolate to roughly  the same value in the
continuum limit, namely $f_{cont} \sim 0.33$. We use shaded bands to mark the
straight lines and their extrapolation to the continuum limit in order to
indicate our estimate in the uncertainty of the analysis. We stress at this
point, that the number $f_{cont} \sim 0.33$ has nothing to do with the number
of colors in SU(3) -- the same value is obtained also for  the case of SU(2). 
 
Let us now try to understand the significance of the value $f_{cont} \sim 0.33$.
For site percolation in three dimensions the critical value for the occupation
probability is $P_c = 0.316$. The fraction $f$ of points available for
clusters in each of the sectors extrapolates in the continuum limit 
to a value of $f_{cont} \sim 0.33$ which is just above the percolation threshold
$P_c = 0.316$ \footnote{As a matter of fact it could be that $f$ tries to
extrapolate to exactly the percolation threshold 0.316, which would be an
extremely beautiful result. However, with the resources currently available to
us we cannot perform a calculation which is accurate enough for a determination
of $f_{cont}$ with sufficient precision to test this hypothesis.}. This
suggests that the clusters indeed grow to infinity in lattice units 
as we approach $a = 0$, and
a continuum limit of the center clusters may be possible. 
We stress again that our calculation should be viewed only as a first 
indication that a continuum limit might be possible and certainly needs to be
substantiated with a (unfortunately very expensive) high precision analysis.

\begin{figure}[t]
\begin{center}
\includegraphics[height=4.45cm,clip]{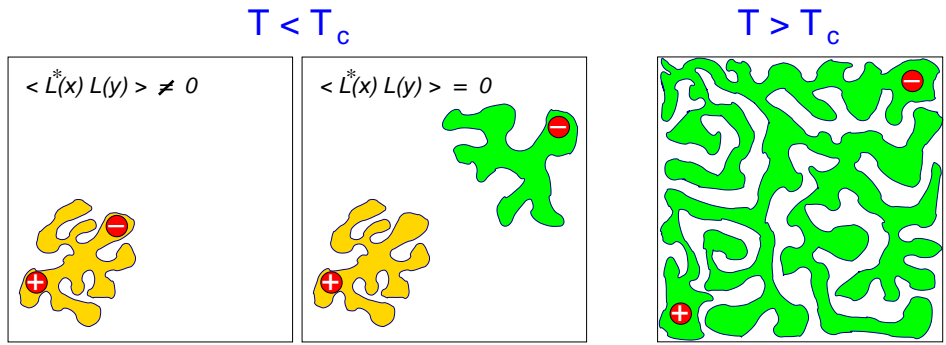}
\end{center}
\caption{Illustration of the confinement mechanism and the deconfinement
transition in terms of center clusters (see the text for explanations). 
\label{mechanism}}
\end{figure}
 
\section{Summary and conclusions}  

In this paper we analyze distribution properties of the local
Polyakov loop in SU(3) and SU(2) lattice gauge theory at finite temperature. 
At each spatial lattice site we identify the nearest center element and study
cluster properties of neighboring sites in the same center sector. While all
center sectors are populated equally below $T_c$, above $T_c$ one of the sectors
is selected spontaneously and becomes more populated at the expense of the
other, subdominant sectors. After introducing the cluster parameter $\delta$
one can construct suitable clusters which start percolating at the
deconfinement transition. 

We address the question whether a continuum limit for the clusters is possible
by using the cluster diameter in physical units as a physical scale. Working on
lattices with different lattice constant $a$ we adjust the cluster parameter
$\delta$ such that the cluster diameter has a fixed value in physical units.
When approaching the continuum limit we find that the fraction of sites which
are available for clusters extrapolates to a value just above the percolation
threshold. This implies that in lattice units the clusters become infinite in
the continuum limit, while in physical units they remain constant (below
$T_c$). This finding indicates that the cluster picture could indeed have a
well defined continuum limit. When analyzing the cluster diameter in physical
units as a function of temperature, we find it is essentially constant below
$T_c$ and becomes infinite above.  

The center clusters and their percolation at the deconfinement temperature give
rise to a simple picture for confinement and the deconfinement transition,
which we illustrate in Fig.~\ref{mechanism}. Below $T_c$ (lhs.\ and center
panels in Fig.~\ref{mechanism}) the center clusters have a characteristic
finite size. If one places two static sources $L(\vec{x})$ and
$L(\vec{y})^\star$ at a distance $|\vec{x} - \vec{y}|$ which is small enough to
fit into one of the clusters (lhs.\ panel), the center phase information is the
same at both positions $\vec{x}$ and $\vec{y}$ and cancels in the correlator
$\langle L(\vec{x}) L(\vec{y})^\star \rangle$ which thus can have a
non-vanishing expectation value. If the distance $|\vec{x} - \vec{y}|$ 
is too large to fit into a single cluster (center panel), then the two sources
will always end up in different clusters. Consequently $L(\vec{x})$ and
$L(\vec{y})^\star$ will be subject to independent fluctuations of the center
phase such that $\langle L(\vec{x}) L(\vec{y})^\star \rangle$ averages to zero.
This averaging does not imply that the correlator vanishes abruptly above a 
fixed value of $|\vec{x} - \vec{y}|$, because the sizes of the clusters 
fluctuate giving rise to the well known exponential decay of 
$\langle L(\vec{x}) L(\vec{y})^\star \rangle$. Above $T_c$ (rhs.\
panel in Fig.~\ref{mechanism}), the center clusters percolate and thus provide
a coherent center information for arbitrary large distances. As a consequence,
above $T_c$ there are non-vanishing contributions to 
$\langle L(\vec{x}) L(\vec{y})^\star \rangle$ at arbitrary large 
distances $|\vec{x} - \vec{y}|$, and the sources are deconfined.

For a further analysis of the center cluster picture several directions need to
be explored. As already discussed, the numerical evidence presented here can be
improved considerably by a large scale precision study of the behavior of the
center clusters using ensembles in a wide range of the lattice constant $a$. In
particular the behavior of the fraction $f$ of available sites for clusters per
center sector should be explored closer to the continuum limit in order to
precisely determine the value it extrapolates to. A second highly important
question is how the center clusters change when full QCD with dynamical
fermions is considered. In this case the deconfinement transition turns into a
crossover and this different behavior should be reflected in the cluster
properties. Preliminary results for that case can be found in \cite{borsanyi}
and a more detailed account on the dynamical case is in preparation.

\vskip5mm
\noindent
{\bf Acknowledgments:} The authors thank Szabolcs Borsanyi, Julia Danzer, 
George Flemming, Zoltan Fodor, Michael Ilgenfritz, Christian Lang and Axel
Maas  for valuable comments. The numerical calculations were  done at the ZID
clusters of the University Graz.

\clearpage

\end{document}